\documentclass[10pt,twocolumn]{article}
\usepackage{latex8}

\usepackage{color,graphics}
\usepackage{url}
\usepackage{epsfig}
\usepackage{amssymb}
\usepackage{amsmath}
\usepackage{amsfonts}

\begingroup
\catcode`\~=11
\gdef\urltilde{\lower 0.6ex\hbox{~}}
\endgroup

 \newcommand{\D}{\mathcal{D}}
\newcommand{\E}{\mathcal{E}} \newcommand{\F}{\mathcal{F}}

 \renewcommand{\L}{\mathcal{L}}
\newcommand{\M}{\mathcal{M}} 
\renewcommand{\O}{\mathcal{O}} \renewcommand{\P}{\mathcal{P}}
 \newcommand{\R}{\mathcal{R}}
 \newcommand{\T}{\mathcal{T}}
 \newcommand{\V}{\mathcal{V}}
\newcommand{\W}{\mathcal{W}}

\newtheorem{theo}{Theorem}

\newtheorem{propo}{Proposition}

\newtheorem{definition}{Definition}

\begin{document}

\title{Probabilistic Logic: Many-valuedness and Intensionality}

\author{Zoran Majki\'c \\
International Society for Research in Science and Technology,\\
 PO Box 2464 Tallahassee, FL 32316 - 2464 USA\\
majk.1234@yahoo.com, ~~~ http://zoranmajkic.webs.com/}


\maketitle \thispagestyle{empty}
%
\begin{abstract}
The probability theory is a well-studied branch of mathematics, in
order to carry out formal reasoning about probability. Thus, it is
important to have a logic, both for computation of probabilities and
for reasoning about probabilities, with a well-defined syntax and
semantics. Both current approaches, based on Nilsson's probability
structures/logics, and on linear inequalities in order to reason
about probabilities, have some weak points.\\
In this paper we have presented the complete revision of both
approaches. We have shown that the full embedding of Nilsson'
probabilistic structure into propositional logic results in a
\emph{truth-functional} many-valued logic, differently from
Nilsson's intuition and current considerations about propositional
probabilistic logic. Than we have shown  that the logic for
reasoning \emph{about} probabilities can be naturally embedded into
a 2-valued intensional FOL with intensional abstraction, by avoiding
current ad-hoc system composed of \emph{two different} 2-valued
logics: one for the classical propositional logic at lower-level,
and a new one at higher-level for  probabilistic constraints with
probabilistic variables. The obtained theoretical results are
applied to Probabilistic Logic Programming.
\end{abstract}



 %
\section{Introduction} \label{section:problogic}
%
The main motivation for an introduction of the intensionality in the
probabilistic-theory of the propositional logic is based on the
desire to have the \emph{full} logical
 embedding  of the probability
into the First-Order Logic (FOL), with a clear difference from the
classic concept of truth of the logic formulae and the concept of
their probabilities. In this way we are able to replace the ad-hoc
syntax and semantics, used in current practice for Probabilistic
Logic Programs \cite{NgSu92,DeDe04,UZVS06,Majk07tp} and
probabilistic deduction \cite{Majk09i}, by the standard syntax and
semantics used for the FOL where the probabilistic-theory properties
are expressed
simply by the particular constraints on their interpretations and models.\\
In this paper we will consider the probabilistic semantics for the
propositional logic (it can be easily extended to predicate logics
 as well) \cite{Nils86,FaHM90} with a fixed
finite set $\Phi = \{p_1,...,p_n\}$ of primitive propositions, which
can be thought of as corresponding to basic probabilistic events.
The set $\L(\Phi)$ of the propositional formulae is the closure of
$\Phi$ under the Boolean operations for conjunction and negation,
$\wedge$ and $\neg$, that is, it is the set of all formulae of the
propositional logic $(\Phi,
\{\wedge, \neg\})$.\\
In order to give the probabilistic semantics to such formulae, we
first need to review briefly the probability theory (see, for
example, \cite{Fell57,Halm50}):
\begin{definition} \label{Def:p-space}
A probability space $(S,\mathcal{X}, \mu)$ consists of a set $S$,
called the sample space, a $\sigma$-algebra $\mathcal{X}$ of subsets
of $S$ (i.e., a set of subsets of $S$ containing $S$ and closed
under complementation and countable union, but not necessarily
consisting of all subsets of $S$) whose elements are called
measurable sets, and a probability measure $\mu:\mathcal{X}
\rightarrow [0,1]$ where
  $[0,1]$ is the closed
interval of reals from 0 to 1. This mapping satisfies
Kolmogorov axioms \cite{Kolm86}:\\
A.1 $~\mu \geq 0$ for all  $X \in \mathcal{X}$.\\
A.2 $~\mu(S) = 1$.\\
A.3 $~\mu(\bigcup_{i\geq 1} X_i) = \sum_{i\geq 1} \mu(X_i)$, $~~~$if
$~X_i$'s are nonempty pairwise
disjoint members of $~\mathcal{X}$.
\end{definition}
The  $\mu(\{s\})$ is the value of probability in a single point of
space $s$.\\
 The property A.3 is called \emph{countable additivity} for
the probabilities in a space $S$. In the case when $~\mathcal{X}$ is
finite set, then we can simplify property A.3 above to \\
A.3' $~\mu(X \bigcup Y) = \mu(X) + \mu(Y)$, \\ if $X$ and $Y$ are
disjoint members of $~\mathcal{X}$, or, equivalently, to the following axiom:\\
A.3'' $~\mu(X ) = \mu(X\bigcap Y) + \mu(X \bigcap \overline{Y})$,\\
where $\overline{Y})$ is the compliment of $Y$ in $S$, so that $\mu(
\overline{X}) = 1 - \mu(X)$.\\
In what follows we will consider only finite sample space $S$, so
that $~\mathcal{X} = \P(S)$ is the set of all subsets of $S$.
 Thus, in our case of a finite set $S$ we obtain, form
A.1 and A.2, that for any $X \in \P(S)$,  $~\mu(X) = \sum_{s \in X}
\mu(\{s\})$.\\
 Based on the work of Nilsson in \cite{Nils86} we can
define for a given propositional logic with a finite set of
primitive proposition $\Phi$ the sample space $S =
\textbf{2}^{\Phi}$, where $\textbf{2} = \{0,1\}\subset [0,1]$, so
that the probability space is equal to the Nilsson structure $N =
(\textbf{2}^{\Phi}, \P(\textbf{2}^{\Phi}),
\mu)$.\\
In his work (page 2, line 4-6 in \cite{Nils86}) Nilsson considered a
Probabilistic Logic "in which the truth values of sentences can
range between $0$ and $1$. The truth value of a sentence in
\emph{probabilistic logic} is taken to be the probability of that
sentence in ordinary first-order logic." That is, he considered this
logic as a kind of a \emph{many-valued}, but not a compositional
truth-valued, logic. But in his paper he did not defined the formal
syntax and semantics for such a probabilistic logic, but only the
matrix equations where the probability of a sentence $\phi \in
\L(\Phi)$ is the sum of the probabilities of the sets of possible
worlds (equal to the set $S = \textbf{2}^{\Phi}$) in which that
sentence is \emph{true}. So that he assigns \emph{two} different
logic values to each sentence $\phi$: one is its probability value
and another is a classic 2-valued truth value in a given possible
world. It is formally contradictory with his intension paraphrased
above. In fact, as we will see in one of the following Section, the
correct formalization of such a many-valued logic \emph{with
probabilistic semantics} is different, and more complex, from his
intuitive initial idea.\\
The \emph{logic} inadequacy of this seminal work \cite{Nils86} of
Nillson is also considered in \cite{FaHa89}, by extending this
Nilsson structure $N = (\textbf{2}^{\Phi}, \P(\textbf{2}^{\Phi}),
\mu)$ into a more general  \emph{probability structure} $M =
(\textbf{2}^{\Phi}, \P(\textbf{2}^{\Phi}), \mu, \pi)$, where $\pi$
associates with each $s \in S = \textbf{2}^{\Phi}$ the truth
assignment $\pi(s):\Phi \rightarrow \textbf{2}$ in the way that we
say that $p \in \Phi$ is \emph{true at} $s$ if $\pi(s)(p) = 1$;
\emph{false at} $s$ otherwise. This mapping $\pi(s)$ can be uniquely
extended to the truth assignment on all formulae in $\L(\Phi)$, by
taking the usual rules of propositional logic (the unique
homomorphic extension to all formulae), and we can associate to each
propositional formula $\phi \in \L(\Phi)$ the set $\phi^M$
consisting of all states $s \in S$ where $\phi$ is true (so that
$~\phi^M = \{s \in S ~|~ \pi(s)(\phi) =
1\}$). In \cite{FaHa89} it was demonstrated that for each Nilsson structure $N$ there is
 an equivalent measurable probability structure $M$, and vice versa.\\
But, differently from Nilsson, they did not define a many-valued
propositional logic, but a kind of 2-valued logic based on
probabilistic constraints. They denoted by $w_N(\phi)$ the
\emph{weight} or \emph{probability} of $\phi$ in Nilsson structure
$N$, correspondent to the value $\mu(\phi^M)$, so that the basic
probabilistic 2-valued constraint can be defined by expressions $c_1
\leq w_N(\phi)$ and $w_N(\phi) \leq c_2$ for given constants
$c_1,c_2 \in [0,1]$. They expected  their logic to be used for reasoning \emph{about} probabilities.\\
But, again, from the logic point of view, they did not defined a
unique logic, but \emph{two different} logics: one for the classical
propositional logic $\L(\phi)$, and a new one for 2-valued
probabilistic constraints obtained from the basic probabilistic
formulae above and Boolean operators $\wedge$ and $\neg$. They did
not consider the introduced symbol $w_N$ as a formal functional
symbol for a mapping $w_N:\L(\Phi) \rightarrow [0,1]$, such that for
any propositional formula $\phi \in \L(\Phi)$, $w_N(\phi) = \mu(\{s
\in S ~|~ \pi(s)(\phi) = 1\})$. Instead of this intuitive meaning
for $w_N$ they considered each expression $w_N(\phi)$ as a
particular probabilistic term (more precisely, as a structured
probabilistic \emph{variable} over the domain of values in
$[0,1]$).\\ It seams that such a dichotomy and difficulty to have
\emph{a unique} 2-valued probabilistic logic, both for an original
propositional formulae in $\L(\Phi)$ and for the probabilistic
constraints, is based on the fact that if we consider $w_N$ as a
function with one argument then it has to be formally represented as
a binary predicate $w_N(\phi,a)$ (for the graph of this function)
where the first argument is a formula and the second is its
resulting probability value. Consequently, a constraint "the
probability of $\phi$ is less or equal to c", has to be formally
expressed by the logic formula $w_N(\phi,a) \wedge \leq (a,c)$ (here
we use a symbol $\leq$ as a built-in rigid binary predicate where
$\leq (a,c)$ is equivalent to $a \leq c$), which is a
\emph{second-order} syntax because $\phi$ is a logic \emph{formula}
in such a unified logic language. That is, the problem of obtaining
the unique logical framework for probabilistic logic comes out with
the necessity of a \emph{reification} feature of this logic
language, analogously to the case of the
intensional semantics for RDF data structures \cite{Majk08ird}.\\
Consequently, we need a logic which is able to deal directly with
reification of logic formulae, and this is the starting point of
this work. In fact as we will see, such an unified logical framework
for the probabilistic theory can be done by a kind of predicate
intensional logics with intensional abstracts, that transforms a
propositional formulae $\phi \in \L(\Phi)$ into an abstracted
\emph{term}, denoted by $\lessdot\phi \gtrdot$. By this approach the
expression $w_N(\lessdot\phi \gtrdot,a) \wedge \leq (a,c)$ remains
to be an ordinary first-order formula. In fact, if $\lessdot\phi
\gtrdot$  is translated into non-sentence "that $\phi$", then the
first-order formula above corresponds to the sentence "the
probability \verb"that" $\phi$ is
true is less than or equal to $c$". \\
More about the intensional FOL can be found in the Appendix. The
decision to define the intensional FOL with abstraction in Appendix,
instead in the first Section is based only on the fact that main
issue of this paper is the probabilistic logic and not intensional
logic.
 But for readers without previous experience about the intensional FOL, we recommend to read this Appendix before the Section 3. Also the particular development
of this intensional FOL is an original contribution of this paper.
 The Plan of this work is the following:\\
In Section 2 we investigate the relationship between Nilsson's
probability structure and many-valued propositional logic. We define
an algebraic probabilistic propositional logic for Nilsson's
structure and show that it is correct truth-functional many-valued
logic for computation of probabilities analogously to Nilsson's
structure.\\
     In Section 3 we define an embedding of the probability theory, both with reasoning about probabilities,
into an intensional FOL with intensional abstraction. Then we show
that the probabilities of propositional formulae corresponds to the
computation of their probabilities in Nilsson's structures, that is,
this intensional FOL is sound and complete w.r.t the measurable
probability structures.\\
     Finally, in Section 4 we apply the theoretical results,
obtained in previous two Sections, to Probabilistic Logic
Programming.
%
 %
\section{Probabilistic logic and many-valuedness} \label{section:problogic-many}
The work of Jan Lukasiewicz was without doubt the most influential
in the development of many-valued modal logics
\cite{Luka30,Luka53,Luka70,MaPr09}. In Lukasiewicz's conception, the
real definition of a logic must be semantic and truth-functional
(the logic connectives are to be truth functions operating on these
logical values): "logic is the science of objects of a specific
kind, namely a science of \emph{logical values}". Many-valued logics
are non-classical logics. They are similar to classical logic
because they accept the principle of truth-functionality, namely,
that the truth of a compound sentence is determined by the truth
values of its component sentences (and so remains unaffected when
one of its component sentences is replaced by another sentence with
the same truth value). But they differ from classical logic by the
fundamental fact that they do not restrict the number of truth
values to only two: they allow for a larger set
 of truth degrees. \\
The Lukasiewicz's work promoted the concept of \emph{logic matrix},
central concept for the construction of many-valued logics, implicit
in the works of C.S.Pierce and E.Post as well. We will briefly
present the previous work for many-valued propositional logics,
based on such matrices:
 The representation
theorems  \cite{Majk06th} for such logics are based on Lindenbaum
algebra of a logic $\L = (\Phi, \O, \Vvdash)$, where $\Phi$ is a set
of propositional variables of a language $\L$, $\O$ is the set of
logical connectives and $\Vvdash$ is the entailment relation of this
logic.\\ We denote by $\L(\Phi)$ the set of all formulae. Lindenbaum
algebra of $\L$ is the quotient algebra $\L(\Phi)/\equiv$, where for
any two formulae $\phi, \psi \in \L(\Phi)$, holds the quivalence
$~~\phi \equiv \psi~~$ iff $~~\phi \Vvdash \psi$ and $\psi \Vvdash
\phi~$. The standard approach to matrices uses a subset $D \subset
A$ of the set of truth values $A$ (nullary operators, i.e., logic
constants),
denominated designated elements; informally a designated element
represent an equivalence class of the \emph{theorems} in $\L$. Given
an algebra $\textbf{A} = (A, \{o\}_{o \in \O})$, the $\O$-matrix is
the pair $(\textbf{A},D)$, where $D \subset A$ is a subset of
designated elements. The \emph{algebraic} satisfaction relation
$\models^a$ ('a' stands for 'algebraic') is defined as follows:
\begin{definition} \label{def:matr} Let $\L = (\Phi, \O,
\Vvdash)$ be a logic, $(\textbf{A},D)$ a $\O$-matrix, and $\phi \in
\L(\Phi)$. Let $v:\Phi \rightarrow A$ be a mapping that assigns the
logic values to propositional variables, and $\overline{v}:\L(\Phi)
\rightarrow A$ be its unique  extension to all formulae of this
logic language $\L(\Phi)$.  We define  the relation $\models^a$
inductively as
follows:\\
1. $(\textbf{A},D)\models^a_v\phi~$ iff $~ \overline{v}(\phi)
\in D$,\\
2. $(\textbf{A},D)\models^a\phi~$ iff $~\overline{v}(\phi) \in D~$
for every $~v:\Phi \rightarrow A$.
\end{definition}
We say that a valuation $\overline{v}$ is \emph{a model} for a
subset of propositional formulae $\Gamma \subset \L(\Phi)$ if for
all formulae $\phi \in
\Gamma$ it holds that $(\textbf{A},D)\models^a_v\phi$.\\
 Any modal 2-valued logic can be considered as
many-valued truth-functional logic for a given set $\W$ of possible
worlds as well, given by a complex algebra $\textbf{A} = (A,
\{\bigcap, \bigcup, /, m, l\})$ of "truth-values", where $A =
\P(\W)$ is the powerset of complete lattice of "truth-values" (empty
set $\emptyset$ is the bottom, while $\W$ is the top "truth-value")
and $\bigcap, \bigcup, /$ are algebraic operations for conjunction,
disjunction and negation (that is,  the set-intersection, set-union
and set-complement operators) respectively, while $m, l$ are
algebraic operations for universal and existential modal logic
operators: in this approach the "truth-value" of a given sentence
$\phi$ is equal to the subset of possible worlds $\|\phi\|$ where
this formula $\phi$ is satisfied, i.e., equal to $\{w \in \W~|~\M
\models_w \phi\} \in \P(\W)$. In such an algebraic logic of a modal
logic, the $\O$-matrix is a pair $(\textbf{A}, D)$ where the set of
designated values is a singleton $D = \{\W\}$. Thus, a sentence
$\phi$ is "true in a Kripke model $\M = (\W,\F,v)$" with a frame
$\F$ of this modal logic iff $(\textbf{A},D)\models^a_v\phi~$, that
is, iff $~ \overline{v}(\phi) \in D$ (i.e., iff $~
\overline{v}(\phi) = \W$), that is, if the "truth-value" of $\phi$
is the top value in $A = \P(\W)$. Notice that instead of this
set-based algebra $\textbf{A} = (A, \{\bigcap, \bigcup, /, m, l\})$
we can use the functional algebra $\textbf{A} = (\widetilde{A},
\{\widetilde{\bigcap}, \widetilde{\bigcup}, \widetilde{/},
\widetilde{m}, \widetilde{l}\})$ with higher order "truth values"
\cite{Majk06FM,Majk06} given by functions in the functional space
$\widetilde{A} = \textbf{2}^{\W}$, where for each $S \in \P(\W)$ we
have the correspondent functional truth-value $f \in
\textbf{2}^{\W}$, such that for any $w \in \W$ it holds that $f(w) =
1$ iff $w \in S$. For example, the function $f \widetilde{\bigcap}
g:\W \rightarrow \textbf{2}$ is defined for any $s \in \W$ by $(f
\widetilde{\bigcap} g)(s) = f(s)\cdot g(s)$, while $(f
\widetilde{\bigcup} g)(s) = f(s)+ g(s)$ and $(\widetilde{/}(f))(s) =
1 - f(s)$.
\\
In what follows, for any function $f \in \textbf{2}^{S}$ we denote
its image by $Im(f) = \{s \in S~ |~f(s) = 1\}$, so that $Im(f
\widetilde{\bigcap} g) = Im(f) \bigcap Im(g)$ and
$im(\widetilde{/}(f)) = S / Im(f)$.
\\
 Here we consider
the possibility to have also infinite matrices for many-valued
truth-functional logics, differently from other approaches that
consider only finite matrices \cite{Bezi04}. In fact, based on
Nilsson probabilistic structure $N = (\textbf{2}^{\Phi},
\P(\textbf{2}^{\Phi}), \mu)$, we can define the following
Probabilistic algebra:
\begin{definition} \label{def:Prob-algebra} Let $N =
(\textbf{2}^{\Phi}, \P(\textbf{2}^{\Phi}), \mu)$ be a Nilsson
structure with a sample space $S = \textbf{2}^{\Phi}$. Then we
define the probabilistic algebra $\textbf{A} = (\textbf{2}^{S}\times
[0,1], \{\curlywedge, \sim\})$ with the $\O$-matrix $(\textbf{A},D)$
where  $D = \{(f, \mu(im(f)))~ |~ f \in \textbf{2}^{S}\}$ is a set
of designated elements, such that the binary operator
"p-conjunction" $\curlywedge$ and unary operator "p-negation" are
defined as follows: for any two $(f,x), (g,y) \in A =
\textbf{2}^{S}\times
[0,1]$,\\
$\curlywedge((f,x),(g,y)) = (f \widetilde{\bigcap} g, \mu(Im(f
\widetilde{\bigcap} g)))$ \\$~~~~~~~~~~~~~~~~~~~~~~~~$ if $~x =
\mu(im(f)), y = \mu(im(f))$;\\ $~~~~~~~~~~~~~~~~~~~~~~(f
\widetilde{\bigcap} g, 0)$ otherwise,\\
 $\sim(f,x) =
(\widetilde{/}(f), \mu(\widetilde{/}(f))$ if $x = \mu(im(f))$;\\
$~~~~~~~~~~~~~~(\widetilde{/}(f), 0)$ otherwise.
\end{definition}
Notice that each truth-value of this algebra is a pair of elements
$a \in A$: first one $\pi_1(a)$ defines the set of possible worlds
in $S$ where a propositional formula in $\L(\Phi)$ is satisfied,
while the second
element $\pi_2(a)$ is the probability of this formula (here $\pi_i, i = 1,2$ are first and second cartesian projections).\\
The set of truth-values in $A$ is infinite, but its subset of
designated elements is finite for the finite set of propositional
variables in $\Phi$.\\
 Let us show that this
algebra represents a \emph{truth-functional} many-valued semantics
for the propositional Nilsson's probabilistic logic.
\begin{propo} \label{propo:Prob-algebra} Let $N =
(\textbf{2}^{\Phi}, \P(\textbf{2}^{\Phi}), \mu)$ be a Nilsson
structure with a sample space $S = \textbf{2}^{\Phi}$, with the
$\O$-matrix $(\textbf{A},D)$ given by Definition
\ref{def:Prob-algebra}, and where $v:\Phi \rightarrow A$ is a
mapping that assigns the logic values to propositional variables,
such that for any $p \in \Phi$, $v(p) = (f, \mu(Im(f)))$ where for
any $s \in S = \textbf{2}^{\Phi}$ it holds that $f(s) = s(p)$. We
denote by $\overline{v}:(\Phi, \{\wedge, \neg\}) \rightarrow
\textbf{A}$ the unique homomorphic extension of $v$ to all formulae
of the propositional logic $(\Phi,
\{\wedge, \neg\})$. \\
Then, for any propositional formula $\phi \in \L(\Phi)$ we have that
$~(\textbf{A},D)\models^a_v\phi~$ implies that
$\pi_2(\overline{v}(\phi))$ is the Nilsson's probability of $\phi$.\\
That is, a many-valued truth-functional assignment $\overline{v}$ is
a \verb"model" for this propositional probabilistic logic.
 \end{propo}
\textbf{Proof:} Let us demonstrate this proposition by structural
induction on formulae $\phi \in \L(\Phi)$:\\
1. For any basic propositional formula $p \in \Phi$ we have that
$~(\textbf{A},D)\models^a_v p~$ (it holds that $v(p) = (f,
\mu(Im(f))) \in D$).\\
Let us suppose that $\phi_1, \phi_2 \in \L(\Phi)$ satisfy this
property, that is, $\overline{v}(\phi_1) = (f, \mu(Im(f))) \in D$
and $\overline{v}(\phi_2) = (g, \mu(Im(g))) \in D$. Then we have the
 following two cases:\\
 2.1 Case when $\phi = \phi_1 \wedge \phi_2$. Then $\overline{v}(\phi) =
\overline{v}(\phi_1 \wedge \phi_2) =\\ =
\curlywedge(\overline{v}(\phi_1),\overline{v}(\phi_2))$ (from the
homomorphic property)\\
$ = \curlywedge((f, \mu(Im(f))),(g, \mu(Im(g))))\\ = (f
\widetilde{\bigcap} g, \mu(Im(f)\bigcap Im(g)))$ (from Definition
\ref{def:Prob-algebra})\\
$ = (f \widetilde{\bigcap} g, \mu(Im(f \widetilde{\bigcap} g))) \in
D$ (from $f \widetilde{\bigcap} g \in
\textbf{2}^{S}$).\\
2.2 Case when $\phi = \neg \phi_1 $. Then $\overline{v}(\phi) =
\overline{v}(\neg \phi_1) =\\ =~ \sim(\overline{v}(\phi_1))$ (from
homomorphic property)\\
$ =~ \sim(f, \mu(Im(f)))\\ = (\widetilde{/}(f), \mu(S /Im(f))$ (from
Definition \ref{def:Prob-algebra})\\
$ = (\widetilde{/}(f), \mu(Im(\widetilde{/}(f))) \in D$ (from $\widetilde{/}(f) \in \textbf{2}^{S}$).\\
Thus, for any $\phi \in \L(\Phi)$ we have that $\overline{v}(\phi) =
(f, \mu(Im(f))$, where $f:S \rightarrow \textbf{2}$ satisfies for
any $s \in S = \textbf{2}^{\Phi}$ that $f(s) = 1$ if
$\overline{s}(\phi) =1$, so that $\mu(Im(f)) =
\pi_2(\overline{v}(\phi))$ is the
Nilsson's probability of the formula $\phi$.\\
From the fact that for any $\phi \in \L(\Phi)$ holds that
$~(\textbf{A},D)\models^a_v\phi~$ we deduce that $\overline{v}$ is a
\emph{model} for this propositional probabilistic logic.
\\$\square$\\
 Notice that for a given propositional logic $(\Phi, \{\wedge,
\neg\})$ and Nilsson's structure $N = (\textbf{2}^{\Phi},
\P(\textbf{2}^{\Phi}), \mu)$, a model $\overline{v}:(\Phi, \{\wedge,
\neg\}) \rightarrow \textbf{A}$ computes the probabilities of all
formulae in $\L(\Phi)$. But there is no way  for this many-valued
propositional logic $(\Phi, \{\wedge, \neg\})$ to reason
\emph{about} these
probabilities.\\
Notice that this approach to probabilistic many-valued logic
demonstrate that, differently from Nilsson's remark (l.3-5, p.72 in
\cite{Nils86}):
\begin{itemize}
  \item "..we present a semantic generalization of ordinary first-order
logic in which the truth values of sentences can range between 0 and
1. The truth-value of a sentence in \emph{probabilistic logic} is
taken to be the \emph{probability} of that sentence .."
\end{itemize}
 the truth-value of a sentence \emph{is not the probability} of that sentence but the
pair $(f,a)$ where the first element $f \in \textbf{2}^{S}$ is a
mapping from the sample space $S = \textbf{2}^{\Phi}$ into the set
$\textbf{2}$;  only its second component $a \in [0,1]$ is the
probability of that sentence. It demonstrates that his intuition was
only god but approximative one, so that  we needed this complete and
formal revision of his original intuition.\\
The second consequence is that, differently from current opinion in
the computer science community that the probabilistic logic is not
truth-functionally many-valued logic, we demonstrated that indeed
\emph{it is}: the truth-value of a complex sentence is functionally
dependent on the truth-values of its proper subsentences, as in
standard many-valued logics.
\section{Probabilistic logic and intensionality} \label{section:problogic-intensional}
%
In order to reason about probabilities of the propositional formulae
we need a kind of 2-valued meta-logic with \emph{reification}
features, thus, a kind of intensional FOL with \emph{intensional
abstraction} presented in the Appendix.\\
Intensional entities are such things as concepts, propositions and
properties.  The \emph{intensional abstracts} are 'that'-clauses.
For example, in the intensional sentence "it is
 necessary that $\phi$", where $\phi$ is a proposition, the 'that $\phi$' is
 denoted by the $\lessdot \phi \gtrdot$, where $\lessdot \gtrdot$ is the intensional abstraction
 operator which transforms a logic formula into a \emph{term}. So
 that the sentence "the probability that A is less then or equal to c" is expressed by
the first-order logic formula $w_N(\lessdot \phi \gtrdot,a) \wedge \leq(a,c)$, where
 $\leq$ is the binary built-in predicate 'is less then or equal', where the usual notation "$a \leq b$" is rewritten
 in this standard predicate-based way by "$\leq(a,b)$", while "the probability that $\phi$ is equal to $a$"
is denoted by the  ground atom $w_N(\lessdot \phi \gtrdot,a)$ with
the  binary "functional" predicate  symbol $w_N$ (in intensional
logic any n-ary function  is represented by the $n+1$-ary predicate
symbol with first $n$ attributes used as arguments of this function
and last $(n+1)$-th attribute for
the resulting function's value, analogously as in FOL with identity).\\
The intensional logic thus is a FOL with terms of FOL and terms
obtained by applying intensional abstraction to the formulae. We
consider a non empty domain $~\D = D_{-1} \bigcup D_I$,  where a
subdomain $D_{-1}$ is made of
 particulars (extensional entities) and contains all real numbers, and the rest $D_I = D_0 \bigcup
 D_1 ...\bigcup D_n ...$ is made of
 universals ($D_0$ for propositions (the 0-ary concepts), and  $D_n, n \geq 1,$ for
 n-ary concepts. So that, similarly to Boolean algebra
for classic logic, we have the Intensional algebra in Definition
\ref{def:int-algebra}  $~Alg_{int} = ~(\D, f, t, Id, Truth,
\{conj_{S}\}_{ S \in \P(\mathbb{N}^2)}, neg,
\\\{exists_{n}\}_{n \in \mathbb{N}})$,  $~~$  with
 binary operations  $~~conj_{S}:D_I\times D_I \rightarrow D_I$,
   unary operation  $~~neg:D_I\rightarrow D_I$,  unary
   operations $~~exists_{n}:D_{I}\rightarrow D_I$.
 The sets $f = \emptyset, ~t = \{<>\}$ are the
empty set and the set with empty tuple $<> \in D_{-1}$ (i.e., the
unique tuple of 0-ary relation) used for logic falsity and truth as
in Codd's relational-database algebra \cite{Codd70}, while $Truth
\in D_0$ is the concept (intension) of the tautology.
 We define that $\D^0 = \{<>\}$, so that $\{f,t\} = \P(\D^0)$.\\
Any extensionalization function $h \in \E$ assigns to the carrier
set $\D_I$ of this algebra (i.e., universals or concepts) an
appropriate extension as follows: for each proposition $u \in D_0$,
$h(u) \in \{f,t\} = \P(\D^0) = \P(\{<>\})  \subset \P(D_{-1})$ is
its
 extension (true or false value), where
$\P$ is the powerset operator, and $h(Truth) = t$; for each n-ary
 concept $u \in D_n$, $h(u)$ is a subset of $\D^n$; in the case of particulars $u \in
 D_{-1}$, we have that $h(u) = u$.
We require that operations $conj, disj$ and $neg$ in this
intensional algebra behave in the expected way with respect to each
extensionalization function (for example, for all $u \in D_0$,
$h(neg(u)) = t$ iff $h(u) = f$, etc..), that is
\begin{center}
 $h = h_{-1}  + \sum_{i\geq 0}h_i:\sum_{i
\geq -1}D_i \longrightarrow D_{-1} +  \sum_{i\geq 0}\P(\D^i)$
\end{center}
 where $h_{-1} = id:D_{-1} \rightarrow D_{-1}$
is the identity, $h_0:D_0 \rightarrow \{f,t\}$ assigns the truth
values in $ \{f,t\}$, to all propositions, and $h_i:D_i \rightarrow
\P(\D^i)$, $i \geq 1$, assigns an extension to all concepts. Thus,
the intensions can be seen as \emph{names} of abstract or concrete
entities, while the extensions correspond to various rules that
these entities play in different worlds: as shown in Appendix by
Bealer-Montague isomorphism, each extensionalization function $h$
can be considered equivalently as a
"possible world" in Montague's semantics for intensional logic. \\
The Tarski-style definitions of truth and validity for this
intensional FOL language $\L$ may be given in the customary way. An
\emph{intensional interpretation} is a mapping between the set $\L$
of formulae of the logic language  and
 intensional entities in $\D$, $I:\L \rightarrow \D$, is a kind of
 "conceptualization", such that  an open-sentence (virtual
 predicate)
 $\phi(x_1,...,x_k) \in \L$ with a tuple of all free variables
 $(x_1,...,x_k)$ is mapped into a k-ary \emph{concept}, that is, an intensional entity  $u =
 I(\phi(x_1,...,x_k)) \in D_k$, and (closed) sentence $\psi$ into a proposition (i.e., \emph{logic} concept) $u =
 I(\psi) \in D_0$ with $I(\top) = Truth \in D_0$ for the FOL tautology $\top$. A language constant $c$ is mapped into a
 particular $\overline{c} = I(c) \in D_{-1}$ if it is a proper name, otherwise in a correspondent concept in $\D$.
The interpretations of singular abstracted terms $\lessdot \phi
\gtrdot_{x_1...x_m}$ will denote an appropriate proposition (for $m
= 0$), property (for $m = 1$), or relation (for $m \geq 2$). Thus,
the mapping of intensional abstracts (terms) into $\D$ will be an
extension of $I$ to all abstract, such that the interpretation of
$\lessdot \phi \gtrdot $
 is equal to the meaning of a proposition $\phi \in \L$, that is, $~~I(\lessdot \phi \gtrdot) =
I(\phi)\in D_0$.  In the case when $\phi$ is an atom
$p_i^m(x_1,..,x_m) \in \L$ then $I(\lessdot
p_i^m(x_1,..,x_m)\gtrdot_{x_1,..,x_m}) = I(p_i^m(x_1,..,x_m)) \in
D_m$.\\
 The basic
intensional logic language $\L_{PR} \subseteq \L$ for probabilistic
theory is composed by propositions in $\L(\Phi)$, with propositional
symbols (0-ary predicate symbols)  $p_i^0 = p_i \in \Phi$ (with
$I(p_i) \in D_0$), with the binary predicate $p_3^2$ for the weight
or probabilistic function $w_N$,  the binary built-in (with constant
fixed extension in any "world' $h \in \E$) predicate $p_2^2$ for
$\leq$ (the binary predicate  $=$ for identity is defined by $ a =
b$ iff $a \leq b$ and $b \leq a$), and two built-in ternary
predicates $p_1^3,p_2^3$, denoted by $\oplus, \odot$, for addition
and multiplication operations $+,\cdot$ respectively as required for
a logic for reasoning about probabilities \cite{FaHM90}. The 0-ary
functional symbols $a,b,c,..$ in this
logic language will be used as  numeric constants for denotation of probabilities in $[0,1]$, i.e., with $I(a) = \overline{a} \in [0,1] \subset D_{-1}$.\\
Consequently, the "worlds" (i.e., the extensionalization functions)
will be reduced to the mappings\\
 $~~h = h_{-1} + h_0  + h_2 + h_3$.\\
We recall that in intensional FOL each n-ary functional symbol is
represented by the (n+1)-ary predicate letter, where the last
attribute is introduced for the resulting values of such a function.
For example, the first attribute of the predicate letter $w_N$ will
contain  the intensional abstract of a propositional formula in
$\L(\Phi)$, while the second place will contain the probabilistic
value in the interval of reals $[0,1] \subset D_{-1}$, so that the
ground atom $w_N(\lessdot \phi \gtrdot,a)$ in $\L_{PR}$ will have
the interpretation $I(w_N(\lessdot \phi \gtrdot,a)) \in D_0$. The
atom $w_N(x,y)$, with variables $x$ and $y$, will satisfy the
functional requirements, that is $I(w_N(x,y)) \in D_2$ with a binary
relation $R = h(I(w_N(x,y))) \in \P(D_0 \times [0,1]) \subseteq
\P(\D^2)$, such that for any $(u,v) \in R$ there is no $v_1 \neq v$
such that $(u,v_1) \in R$. Obviously, for this intensional logic we
have that $h(I(w_N(\lessdot \phi \gtrdot,a))) = t$ iff $(I(\lessdot
\phi \gtrdot),I(a)) = (I(\phi), \overline{a})\in R$.
\\
Analogously, for the ground atom $\oplus(a,b,c)$, with $\overline{a}
= I(a), \overline{b} = I(b), \overline{c} = I(c) \in  D_{-1}$ real
numbers, we have that $I(\oplus(a,b,c)) \in D_0$ such that for any
"world" $h \in \E$ we have that $h(I(\oplus(a,b,c))) = t$ iff
$\overline{a} + \overline{b} = \overline{c}$ (remember that for
elements $\overline{a} = I(a) \in D_{-1}$ we have that $h(I(a)) =
\overline{a}$). For addition of more than two elements in this
intensional logic we will use intensional abstract, for example for
the sum  of three elements we can use a ground formula
$\oplus(a,b,d) \wedge \oplus(d,c,e)$, such that it holds that
$h(I(\oplus(a,b,d) \wedge \oplus(d,b,c))) =
conj(I(\oplus(a,b,d)),I(\oplus(d,c,e))) = t$ iff $\overline{a} +
\overline{b} = \overline{d}$ and  $\overline{d} + \overline{c} =
\overline{e}$, that is, iff $\overline{a} + \overline{b} +
\overline{c} = \overline{e}$. The fixed extensions of the two
built-in ternary predicates $\oplus(x,y,z)$ and $\odot(x,y,z)$ are
equal to:\\
$R_{\oplus} = h(I(\oplus(x,y,z))) = \{(u_1,u_2,u_1 + u_2)~|~u_1,u_2
\in
 D_{-1}$ are real numbers $ \}$,\\
$R_{\odot} = h(I(\odot(x,y,z))) = \{(u_1,u_2,u_1 \cdot
u_2)~|~u_1,u_2 \in
  D_{-1}$ are real numbers $ \}$.\\
The built-in binary predicate $\leq$ satisfies the following
requirements for its intensional interpretation: $I(\leq(x,y)) \in
D_2$ such that for every $h \in \E$ it holds that its fixed
extension is a binary relation $R_{\leq} = h(I(\leq(x,y))) =
\{(u,v)~|~u,v \in  D_{-1}$ are real numbers and $u \leq v \}$, with
the property that $h(I(\leq(a_1,a_2))) = t$ iff $(I(a_1),I(a_2)) =
(\overline{a}_1, \overline{a}_2) \in R_{\leq}$.
\begin{definition}  \label{def:Probab-intlogic}
Intensional FOL $\L_{PR}$ is a probabilistic logic with a
\emph{probability structure} $M = (\textbf{2}^{\Phi},
\P(\textbf{2}^{\Phi}), \mu, \pi)$ if its intentional interpretations
satisfy the following property for any propositional formula $\phi
\in \L(\Phi) \subseteq \L_{PR}$: $~~h(I(w_N(\lessdot \phi
\gtrdot,a))) = t$\\ iff $~I(a) = \sum_{s \in \textbf{2}^{\Phi} ~\&~
\pi(s)(\phi) = 1} ~\mu(\{s\})$.
 \end{definition}
 Let us show that the binary predicate $w_N$ is a functional
 built-in predicate, whose extension is equal in every possible
 "world" $h \in \E$, and that the probability structure can use $\E$
 as the set of possible worlds in the place  of Nilsson's set
 $\textbf{2}^{\Phi}$. That is, we can replace Nilson's structure with the intensional probability structure
 $M_I = (\E, \P(\E), \mu, \pi)$.
\begin{propo}  \label{prop:Probab-intlogic}
Intensional FOL $\L_{PR}$ is a probabilistic logic with a
\emph{probability structure} $M = (\textbf{2}^{\Phi},
\P(\textbf{2}^{\Phi}), \mu, \pi)$ if  $w_N$ is a built-in functional
symbol such that its fixed extension is equal to:\\
$R_{w_N} = h(I(w_N(x,y))) = \{(I(\phi), I(a)) ~|~\phi \in \L(\Phi)$
and $I(a) = \sum_{h_1 \in \E ~\&~ h_1(I(w_N(\lessdot \phi
\gtrdot,a))) = t} ~\mu(\{is^{-1}(h_1)\})\}$\\ where the mapping
$is:\textbf{2}^{\Phi}\rightarrow \E$ is a bijection, and $is^{-1}$
its inverse.
 \end{propo}
 \textbf{Proof:}
Let us show that
 there is a bijection $is:\textbf{2}^{\Phi}\rightarrow \E$ between the sets $\textbf{2}^{\Phi}$ and $\E$.
 In fact, let $v \in \textbf{2}^{\Phi}$ be extended (in the unique standard homomorphic way) to all
 propositional formulae by $\overline{v}:\L(\Phi) \rightarrow \textbf{2}$. This propositional valuation corresponds to the intensional
 interpretation $(I,h)$ obtained, for any sentence $\phi \in
\L(\Phi)$, by $h(I(\phi)) = is_\textbf{2}(\overline{v}(\phi))$,
where $is_\textbf{2}:\textbf{2}\rightarrow \{f,t\}$ is a bijection
of these two lattices such that $is_\textbf{2}(0) = f,
~is_\textbf{2}(1) = t$. We have seen that all predicate symbols with
arity bigger than $0$ of our intensional probabilistic logic
$\L_{PR}$ are \emph{built-in predicates} (that do not depend on $h
\in \E$), so that for a fixed intensional interpretation $I$, any
two extensionalization functions $h, h'$ differ only on propositions
in $D_0$, so that we obtain the bijective mapping $is: v \mapsto h$,
such that $v = is_\textbf{2}^{-1}\circ h \circ I$, where $\circ$
denotes the composition of functions.\\
From Tarski's constraint (T) of intensional algebra (in Appendix) we
have that for any ground
atom $w_N(\lessdot \phi \gtrdot,a)$ it holds that:\\
$I(w_N(\lessdot \phi \gtrdot,a))
 = t$ iff $(I(\phi), I(a)) \in
h(I(w_N(x,y)))$. Thus, form Definition \ref{def:Probab-intlogic}, we
obtain that $(I(\phi), I(a)) \in h(I(w_N(x,y)))$ iff $I(a) = \sum_{s
\in \textbf{2}^{\Phi} ~\&~ \pi(s)(\phi) = 1} ~\mu(\{s\})$.\\
Consequently,\\
$R_{w_N} = h(I(w_N(x,y))) = \{(I(\phi), I(a)) ~|~\phi \in \L(\Phi)$
and $I(a) = \sum_{s \in \textbf{2}^{\Phi} ~\&~ \pi(s)(\phi) = 1}
~\mu(\{s\})\}$, where $I(w_N(x,y)) \in D_2$, $I(\phi) \in D_0$ and $I(a) \in [0,1] \subseteq D_{-1}$.\\
 But from a bijection $is$, instead of $s \in
\textbf{2}^{\Phi}$ we can take $h_1 = is(s) \in \E$, and the fact $~
\pi(s)(\phi) = 1$ means that "$\phi$ is true in the state $s$" can
be equivalently replaced by "$\phi$ is true in the "world" $h_1 =
is(s)$", that is by condition $~ h_1(I(w_N(\lessdot \phi
\gtrdot,a))) = t$, so that definition for the extension of the
binary relation $R_{w_N}$ for Nilsson's probabilities of
propositional formulae, given in this proposition, is correct. This
extension is constant in any "possible world" in $\E$, thus, the
binary functional-predicate $w_N$ is a built-in predicate in this
intensional FOL $\L_{PR}$.
\\$\square$\\
Consequently, the sentence "the probability that $\phi$ is equal to
a", expressed by the ground atom $w_N(\lessdot \phi \gtrdot, a)$, is
true iff $~~h(I(w_N(\lessdot \phi \gtrdot, a))) =
  t~~$ iff \\$~~(I( \phi),
\overline{a}) \in h(I(w_n(x,y)) = R_{w_N}$.\\
Thus, for the most simple linear inequality, "the probability that
$\phi$ is less then or equal to c", expressed by the  formula
$\exists x(w_N(\lessdot \phi \gtrdot, x) \wedge \leq(x,c))$, is true iff $~~h(I(\exists x(w_N(\lessdot \phi \gtrdot, x) \wedge \leq(x,c)))) = t$\\
iff $~~(u,I(c)) \in R_{\leq}$ where $u \in [0,1] \subset D_{-1}$ is determined by $(v,u) \in R_{w_N}$ where $v = I( \phi)\in D_0$.\\
 Analogously to the results obtained for a logic for reasoning
about probabilities in \cite{FaHM90}, we obtain the following
property:
\begin{theo}  \label{prop:Probab-complete} The intensional FOL
$\L_{PR}$ with built-in binary predicate $w_N$ defined in
Proposition \ref{prop:Probab-intlogic}, built-in binary predicate
$\leq$ and ternary built-in predicates $\oplus, \odot$, is sound and
complete with respect to the measurable probability structures.
\end{theo}
\textbf{Proof:} We will follow the demonstration analogous to the
demonstration of Theorem 2.2 in \cite{FaHM90} for the sound and
complete axiomatization  of the axiomatic system $AX_{MEAS}$ for
logic reasoning about probabilities, divided into three parts, which
deal respectively with propositional reasoning, reasoning about
linear inequalities, and reasoning about probabilities:\\
1. Propositional reasoning: set of all instances of propositional
tautologies, with unique inference rule Modus Ponens.\\
2. Reasoning about linear inequalities: set of all instances of
valid formulae about linear inequalities of the form $a_1\cdot x_1
+...+ a_k\cdot x_k \leq c$, where $a1,...,a_k$ and $c$ are integers
with $k \geq 1$, while $x1,...,x_k$ are probabilistic variables.\\
3. Reasoning about probability function:\\
    3.1 $w(\phi) \geq 0$ (nonnegativity)\\
    3.2 $w(true) = 1$ (the probability of the event true is 1)\\
    3.3 $w(\phi \wedge \psi) + w(\phi \wedge \neg \psi) = w(\phi)$
(additivity)\\
    3.4 $w(\phi) = w(\psi)$ if $\phi \equiv \psi$ (distributivity).\\
It is easy to verify that for any propositional axiom $\phi$, we
have that for all "worlds" $h \in \E$ it holds that $h(I(\phi)) =
t$, so that it is true in the S5 Kripke model of the intensional FOL
given in Definition \ref{def:Semant} (Appendix), because all
algebraic operations in $~Alg_{int}$ in Definition
 \ref{def:int-algebra}(Appendix) are defined in order to satisfy
standard propositional logic. Moreover, the Modus Ponens rule is
satisfied in every "world" $h \in \E$. Thus, the point 1 above is
satisfied by intensional logic $\L_{PR}$.\\
The definition of built-in predicates $\odot, \oplus$ and $\leq$
satisfy all linear inequalities, thus the point 2 above.\\
The definition of binary predicate $w_N(x,y)$ is given in order to
satisfy Nilsson's probability structure, thus all properties of
probability funcion in the point 3 above are satisfied by $w_N(x,y)$
built-in predicate in every "world" $h \in \E$. Consequently, the
soundness and completeness of the intensional logic $\L_{PR}$  with
respect to measurable probability structures, based on the Theorem
Theorem 2.2 in \cite{FaHM90} is satisfied.
\\$\square$\\
For example, the satisfaction of the linear inequality $a_1\cdot x_1
+ a_2\cdot x_2 \leq c$, where $x_1$ and $x_2$ are the probabilities
of the propositional formulae $\phi_1$ and $\phi_2$ relatively,
 (here the list of quantifications $(\exists x_1)...(\exists x_k)$ is abbreviated by
$(\exists x_1,...,x_k)$),  expressed by the following intensional  formula,\\
$(\exists x_1,x_2,y_1,y_2,y_3)(w_N(\lessdot \phi_1 \gtrdot, x_1)
\wedge w_N(\lessdot \phi_2 \gtrdot, x_2) \wedge \odot(a_1,x_1,y_1)
\wedge \odot(a_2,x_2,y_2) \wedge \oplus(y_1,y_2,y_3) \wedge
\leq(y_3,c))$, is true iff $~~(I( \phi_1), u_1), (I( \phi_2), u_2)
\in R_{w_N}$, $(I(a_1),u_1,v_1), (I(a_2),u_2,v_2) \in R_{\odot}$,
$(v_1,v_2,v_3)) \in R_{\oplus}$
 and $(v_3,I(c)) \in
R_{\leq}$, \\where $u_1,u_2,v_1,..,v_3 \in  D_{-1}$ are real numbers.\\
Analogously, the satisfaction of any linear inequality $a_1\cdot x_1
+...+ a_k\cdot x_k \leq c$, where $x_i$ are the probabilities of the
propositional formulae $\phi_i$ for  $i = 1,...,k$, $k \geq 2$, can
be expressed by the logic formula\\
 $(\exists x_1,...,x_k,y_1,...,y_k,z_1,...,z_k)(w_N(\lessdot \phi_1 \gtrdot,
x_1) \wedge...\\ \wedge w_N(\lessdot \phi_k \gtrdot, x_k) \wedge
\odot(a_1,x_1,y_1) \wedge... \wedge \odot(a_k,x_k,y_k)\\  \wedge
\oplus(0,y_1,z_1) \wedge \oplus(z_1,y_2,z_2)\wedge...\wedge
\oplus(z_{k-1},y_k,z_k) \\ \wedge \leq(z_k,c))$.\\
Based on the  Theorem 2.9 in \cite{FaHM90} we can conclude that the
problem of deciding whether such an intensional formula in $\L_{PR}$
is satisfiable in a measurable probability structure of Nilsson is
NP-complete.
\section{Application to Probabilistic Logic Programs}
The semantics of the interval-based Probabilistic Logic Programs,
based on possible worlds with the fixpoint semantics for such
programs \cite{NgSu92}, has been considered valid for more than 13
years. But some years ago, when I worked with Prof.Subrahmanian
director of the UMIACS institute, I have had the opportunity to
consider the general problems of (temporal) probabilistic databases
\cite{UZVS05}, to analize their semantics of interval-based
Probabilistic Logic Programs,  and to realize that unfortunately it
was not correctly defined.\\
Because of that I formally developed,  in \cite{Majk05E}, the
reduction of (temporal) probabilistic databases into Constraint
Logic Programs. Consequently,  it was possible to apply the interval
PSAT in order o find the models of such interval-based probabilistic
programs, as presented  and compared with other approaches in
\cite{Majk07tp}. Moreover, in this complete revision it was
demonstrated that the Temporal-Probabilistic Logic Programs can be
reduced into the particular case of the ordinary
Probabilistic Logic Programs, so that our application of intensional semantics  we can apply only to this last general case of Logic Programs.\\
In what follows I will briefly introduce the syntax of Probabilistic
Logic Programs. More about it can be found in the original work in
\cite{NgSu92} and in its last revision in \cite{Majk07tp}. Let
$ground(P)$ denote the set of all ground instances of rules of a
Probabilistic Logic Program $P$ with a given domain for object
variables, and let $H$ denote the Herbrand base of this program $P$.
Then, each ground instance of rules in $ground(P)$
has the following syntax:\\
$(*)~~~~~~~~~A:\mu_0 \leftarrow \phi_1:\mu_1 \wedge...\wedge
 \phi_m:\mu_m$,\\
where $A \in H$ is a ground atom in a Herebrand base $H$, $\phi_i, i
\geq 1$ are logic formulae composed by ground atoms and standard
logic connectives $\wedge$ and $\neg$, while $\mu_i = (b_i, c_i),i
\geq 0$, where $b_i,c_i \in [0,1]$, are the lower and upper
probability
boundaries.\\
The expression $~\phi_i:\mu_i~$ is a probabilistic-annotated
(p-annotated) basic formula, which is true if the probability $x_i$
of the ground formula $\phi_i$ is between $b_i$ and $c_i$; false
otherwise. Thus, this basic p-annotated formula is the particular
case of the 2-valued
probabilistic formula:\\
 $(**)~~~~~~~~~(1\cdot x_i \geq a_i) \wedge (1 \cdot x_i \leq
b_i)$\\ composed
by two linear inequalities.\\
Consequently, the standard logic embedding of annotated
interval-based logic programs can be easily obtained by the
intensional logic $\L_{PR}$ described in Section
\ref{section:problogic-intensional} where $\Phi$ is equal to the
Herbrand base $H$ of the annotated interval-based probabilistic
logic program $P$.\\
Thus, based on the translation $(**)$, the logic formula in
intensional logic $\L_{PR}$ correspondent to basic annotated formula
$~\phi_i:\mu_i~$ of the annotated logic program $ground(P)$, is equal to the following first-order closed formulae with  a variable $x_i$:\\
$~~~~~~~~~\exists x_i(w_N(\lessdot \phi_i\gtrdot, x_i) \wedge
\leq(b_i,x_i)
\wedge \leq(x_i,c_i))$.\\
Based on this translation, the rule $(*)$ of the annotated logic
Program $ground(P)$ can be replaced by the following rule of an
intensional probabilistic
logic program:\\
$\exists x_0(w_N(\lessdot A\gtrdot, x_0) \wedge \leq(b_0,x_0) \wedge
\leq(x_0,c_0))~~ \leftarrow \\
~~~~\leftarrow ~~ \exists x_1(w_N(\lessdot \phi_1\gtrdot, x_1)
\wedge \leq(b_1,x_1) \wedge \leq(x_1,c_1)) \wedge ...\\
~~~~~~~\wedge ~\exists x_m(w_N(\lessdot \phi_m\gtrdot, x_m) \wedge
\leq(b_m,x_m) \wedge
\leq(x_m,c_m))$,\\
with  the variables $x_0, x_1,...,x_m$.\\
In this way we obtain a grounded intensional probabilistic logic
program $P_{PR}$, which has both the syntax and semantics different
from the original annotated probabilistic logic program $ground(P)$.\\
As an alternative to this full intensional embedding of the
annotated logic programs into the first-order intensional logic, we
can use a partial embedding by preserving the old ad hoc annotated
syntax of the probabilistic program $ground(P)$, by extending the
standard predicate-based syntax of the intensional FOL logic with
annotated formulae, and by  defining only the new intensional
interpretation $I$ for these annotated formulae, as follows:
 $~~~I(\phi_i:\mu_i) = \\ = I((\exists x)(w_N(\lessdot
\phi_i\gtrdot, x))\wedge (\leq(b_i,x))\wedge \leq(x,c_i)))\\=
exist(conj_{\{(1,1)\}}(I(w_N(\lessdot \phi_i\gtrdot,
x)), conj_{\{(1,1)\}}(I(\leq(b_i,x)),I(\leq(x,c_i)))))$,\\
where $I(w_N(\lessdot \phi_i\gtrdot,
x)),I(\leq(b_i,x)),I(\leq(x,c_i)) \in D_1$.\\
 So that $h(I(\phi_i:\mu_i)) = t~~$ iff $~~h(exist(u_1)) = t~~$,
where $u_1 = conj_{\{(1,1)\}}(I(w_N(\lessdot \phi_i\gtrdot, x)),
conj_{\{(1,1)\}}(I(\leq(b_i,x)),I(\leq(x,c_i)))) \in D_1$,\\ $~~$
iff $~~\exists u (u \in h(u_1))~~$
 iff $~~(I(b_i),u),
(u,I(c_i)) \in R_{\leq}$, where $u \in [0,1] \subset D_{-1}$ is a
particular assignment  for a variable $x$ determined by $~(v,u) \in
R_{w_N}$ where $v = I(\phi_i)$. \\
Notice that,  from Appendix and the fact that $u_1 \in D_1$,
$~~h(exist(u_1))= f_{<>}(h(I(w_N(\lessdot \phi_i\gtrdot, x))
\bowtie_{\{(1,1)\}} (h(I(\leq(b_i,x))) \bowtie_{\{(1,1)\}}
h(I(\leq(x,c_i))))\\ = f_{<>}(h(I(w_N(\lessdot \phi_i\gtrdot,
x))\bigcap h(I(\leq(b_i,x))) \bigcap
h(I(\leq(x,c_i)))$, where $f_{<>}(R) = f$ if $R = \emptyset$; $t$ otherwise.\\
 The advantage of this second, partial
embedding is that we can preserve the  old syntax for (temporal)
probabilistic logic programs \cite{NgSu92,Majk07tp}, but providing
to them the standard intensional FOL semantics instead of current ad
hoc semantics for such a kind of logic programs.
\section{Conclusion}
There are two consequences of this full and natural embedding of the
probability theory in a
 logic:
\begin{enumerate}
  \item The full embedding of Nilsson' probabilistic structure into
propositional logic results in a \emph{truth-functional} many-valued
logic, differently from Nilsson's intuition and current
considerations about propositional probabilistic logic.
  \item The logic for reasoning \emph{about} probabilities can be embedded into an intensional
FOL that  remains to be \emph{2-valued logic}, both for
propositional formulae in $\L(\Phi)$ and predicate formulae for
probability constraints, based on the binary built-in predicate
$\leq$ and binary predicate $w_N$ used for the probability function,
where the basic propositional letters in $\Phi$ are formally
considered as nullary predicate symbols.
\end{enumerate}
 The intensional FOL for reasoning about probabilities is obtained by the particular fusion
of the intensional algebra (analogously to Bealer's approach) and
Montague's possible-worlds modal logic for the semantics of the
natural language. In this paper we enriched such a logic framework
by a number of built-in binary and ternary predicates, which can be
used to define the basic set of probability inequalities and to
render the probability weight function $w_N$ an explicit object in
this logic language. We conclude that this intensional FOL logic
with intensional abstraction is a good candidate language for
specification of Probabilistic Logic Programs, and we apply two
different approaches: first one is obtained by the translation of
the annotated syntax of current logic programs into this intensional
FOL; the second one, instead, modify only the semantics of these
logic programs by preserving their current ad-hoc annotated syntax.


\bibliographystyle{IEEEbib}
\bibliography{medium-string,krdb,mydb}

\section{Appendix: Intensional FOL language with intensional abstraction} \label{section:intensional}
%
 Intensional entities are such concepts as
propositions and properties. What make them 'intensional' is that
they violate the principle of extensionality; the principle that
extensional equivalence implies identity. All (or most) of these
intensional entities have been classified at one time or another as
kinds of Universals \cite{Beal93}.\\
We consider a non empty domain $~\D = D_{-1} \bigcup D_I$,  where a
subdomain $D_{-1}$ is made of
 particulars (extensional entities), and the rest $D_I = D_0 \bigcup
 D_1 ...\bigcup D_n ...$ is made of
 universals ($D_0$ for propositions (the 0-ary concepts), and  $D_n, n \geq 1,$ for
 n-ary concepts.\\
 The fundamental entities
are \emph{intensional abstracts} or so called, 'that'-clauses. We
assume that they are singular terms; Intensional expressions like
'believe', mean', 'assert', 'know',
 are standard two-place predicates  that take 'that'-clauses as
 arguments. Expressions like 'is necessary', 'is true', and 'is
 possible' are one-place predicates that take 'that'-clauses as
 arguments. For example, in the intensional sentence "it is
 necessary that $\phi$", where $\phi$ is a proposition, the 'that $\phi$' is
 denoted by the $\lessdot \phi \gtrdot$, where $\lessdot \gtrdot$ is the intensional abstraction
 operator which transforms a logic formula into a \emph{term}. Or, for example, "x believes that $\phi$" is given by formula
$p_i^2(x,\lessdot \phi \gtrdot)$ ( $p_i^2$ is binary 'believe'
predicate).\\
Here we will present an  intensional FOL  with slightly different
intensional abstraction than that originally presented  in
\cite{Beal79}:
 \begin{definition} \label{def:bealer}
  The syntax of the First-order Logic language with intensional abstraction
$\lessdot \gtrdot$, denoted by $\L$, is as follows:\\
 Logic operators $(\wedge, \neg, \exists)$; Predicate letters in $P$
 (functional letters are considered as particular case of predicate
 letters); Variables $x,y,z,..$ in $\V$; Abstraction $\lessdot \_ \gtrdot$, and punctuation
 symbols (comma, parenthesis).
 With the following simultaneous inductive definition of \emph{term} and
 \emph{formula}   :\\
   1. All variables and constants (0-ary functional letters in P) are terms.\\
   2. If $~t_1,...,t_k$ are terms, then $p_i^k(t_1,...,t_k)$ is a formula
 ($p_i^k \in P$ is a k-ary predicate letter).\\
   3. If $\phi$ and $\psi$ are formulae, then $(\phi \wedge \psi)$, $\neg \phi$, and
 $(\exists x)\phi$ are formulae. \\
   4. If $\phi(\textbf{x})$ is a formula (virtual predicate) with a list of free variables in $\textbf{x} =(x_1,...,x_n)$ (with ordering
from-left-to-right of their appearance in $\phi$), and  $\alpha$ is
its sublist of \emph{distinct} variables,
 then $\lessdot \phi \gtrdot_{\alpha}^{\beta}$ is a term, where $\beta$ is the remaining list of free variables preserving ordering in $\textbf{x}$ as well. The externally quantifiable variables are the \emph{free} variables not in $\alpha$.
  When $n =0,~ \lessdot \phi \gtrdot$ is a term which denotes a
proposition, for $n \geq 1$ it denotes
 a n-ary concept.\\
An occurrence of a variable $x_i$ in a formula (or a term) is
\emph{bound} (\emph{free}) iff it lies (does not lie) within a
formula of the form $(\exists x_i)\phi$ (or a term of the form
$\lessdot \phi \gtrdot_{\alpha}^{\beta}$ with $x_i \in \alpha$). A
variable is free (bound) in a formula (or term) iff it has (does not
have) a free occurrence in that formula (or term).\\ A
\emph{sentence} is a formula having no free variables. The binary
predicate letter $p_1^2$ for identity is singled out as a
distinguished logical predicate and formulae of the form
$p^2_1(t_1,t_2)$ are to be rewritten in the form $t_1 \doteq t_2$.
We denote by $R_{=}$ the binary relation obtained by standard
Tarski's interpretation of this predicate $p^2_1$. The logic
operators $\forall, \vee, \Rightarrow$ are defined in terms of
$(\wedge, \neg, \exists)$ in the usual way.
\end{definition}
The \emph{intensional interpretation} of this intensional FOL is a
mapping between the set $\L$ of formulae of the logic language  and
 intensional entities in $\D$, $I:\L \rightarrow \D$, is a kind of
 "conceptualization", such that  an open-sentence (virtual
 predicate)
 $\phi(x_1,...,x_k)$ with a tuple of all free variables
 $(x_1,...,x_k)$ is mapped into a k-ary \emph{concept}, that is, an intensional entity  $u =
 I(\phi(x_1,...,x_k)) \in D_k$, and (closed) sentence $\psi$ into a proposition (i.e., \emph{logic} concept) $v =
 I(\psi) \in D_0$ with $I(\top) = Truth \in D_0$ for the FOL tautology $\top$. A language constant $c$ is mapped into a
 particular $a = I(c) \in D_{-1}$ if it is a proper name, otherwise in a correspondent concept in
$\D$. \\
 An assignment $g:\V \rightarrow \D$ for variables in $\V$ is
applied only to free variables in terms and formulae.  Such an
assignment $g \in \D^{\V}$ can be recursively uniquely extended into
the assignment $g^*:\T \rightarrow \D$, where $\T$ denotes the set
of all terms (here $I$ is an intensional interpretation of this FOL,
as explained
in what follows), by :\\
1. $g^*(t) = g(x) \in \D$ if the term $t$ is a variable $x \in
\V$.\\
2. $g^*(t) = I(c) \in \D$ if the term $t$ is a constant $c \in
P$.\\
3. if $t$ is an abstracted term $\lessdot \phi
\gtrdot_{\alpha}^{\beta}$,  then $g^*(\lessdot \phi
\gtrdot_{\alpha}^{\beta}) = I(\phi[\beta /g(\beta)] ) \in D_k, k =
|\alpha|$ (i.e., the number of variables in $\alpha$), where
$g(\beta) = g(y_1,..,y_m) = (g(y_1),...,g(y_m))$ and $[\beta
/g(\beta)]$ is a uniform replacement of each i-th variable in the
list $\beta$
with the i-th constant in the list $g(\beta)$. Notice that $\alpha$ is the list of all free variables in the formula $\phi[\beta /g(\beta)]$.\\
  We denote by $~t/g~$ (or $\phi/g$) the ground term (or
formula) without free variables, obtained by assignment $g$ from a
term $t$ (or a formula $\phi$), and by  $\phi[x/t]$ the formula
obtained by  uniformly replacing $x$ by a term $t$ in $\phi$.\\
The distinction between intensions and extensions is important
 especially because we are now able to have and \emph{equational
 theory} over intensional entities (as  $\lessdot \phi \gtrdot$), that
 is predicate and function "names", that is separate from the
 extensional equality of relations and functions.
 An extensionalization function $h$ assigns to the intensional elements of $\D$ an appropriate
extension as follows: for each proposition $u \in D_0$, $h(u) \in
 \{f,t\} \subseteq \P(D_{-1})$ is its
 extension (true or false value); for each n-ary
 concept $u \in D_n$, $h(u)$ is a subset of $\D^n$
 (n-th Cartesian product of $\D$); in the case of particulars $u \in
 D_{-1}$, $h(u) = u$.\\
The sets $f, t$
   are empty set $\{\}$ and set $\{<>\}$ (with the empty tuple $<> \in D_{-1}$ i.e. the unique tuple of 0-ary relation)
 which may be thought of
 as falsity and truth, as those used  in the Codd's relational-database algebra \cite{Codd70} respectively,
 while $Truth \in D_0$ is the concept (intension)
of the tautology. \\
 We define that $\D^0 = \{<>\}$, so that $\{f,t\} = \P(\D^0)$.
 Thus we have:
 \\$h = h_{-1}  + \sum_{i\geq 0}h_i:\sum_{i
\geq -1}D_i \longrightarrow D_{-1} +  \sum_{i\geq 0}\P(D^i)$,\\
 where $h_{-1} = id:D_{-1} \rightarrow D_{-1}$
is identity, $h_0:D_0 \rightarrow \{f,t\}$ assigns truth values in $
\{f,t\}$, to all propositions, and $h_i:D_i \rightarrow \P(D^i)$,
$i\geq 1$, assigns extension to all concepts, where $\P$ is the
powerset operator. Thus, intensions can be seen as \emph{names} of
abstract or concrete entities, while extensions correspond to
various rules that these entities play in different worlds.\\
\textbf{Remark:} (Tarski's constraint) This semantics has to
preserve Tarski's semantics of the FOL, that is, for any formula
$\phi \in \L$ with the tuple of free variables $(x_1,...,x_k)$, any
assignment $g \in \D^{\V}$, and every $h \in \E$ it has to be
satisfied that: \\ (T)$~~~h(I(\phi/g)) = t~~$ iff
$~~(g(x_1),...,g(x_k)) \in h(I(\phi))$.\\
$\square$\\
 Thus, intensional  FOL  has the simple Tarski
first-order semantics, with a decidable
 unification problem, but we need also the actual world mapping
 which maps any intensional entity to its \emph{actual world
 extension}. In what follows we will identify a \emph{possible world} by a
 particular mapping which assigns to intensional entities their
 extensions in such possible world. That is direct bridge between
 intensional FOL  and possible worlds representation
 \cite{Lewi86,Stal84,Mont70,Mont73,Mont74}, where intension of a proposition is a
 \emph{function} from possible worlds $\widetilde{\W}$ to truth-values, and
 properties and functions from $\widetilde{\W}$ to sets of possible (usually
 not-actual) objects.\\
 Here $\E$ denotes the set of possible
\emph{extensionalization functions} that satisfy the constraint (T);
they can be considered as \emph{possible worlds} (as in Montague's
intensional semantics for natural language \cite{Mont70,Mont74}), as
demonstrated in \cite{Majk08in,Majk08ird}, given by the bijection
$~~~is:\W \simeq \E$.\\
Now we are able to define formally this intensional semantics:
 \begin{definition} \label{def:intensemant} \textsc{Two-step \textsc{I}ntensional \textsc{S}emantics:}
Let $\mathfrak{R} = \bigcup_{k \in \mathbb{N}} \P(\D^k) = \sum_{k\in
\mathbb{N}}\P(D^k)$ be the set of all k-ary relations, where $k \in
\mathbb{N} = \{0,1,2,...\}$. Notice that $\{f,t\} = \P(\D^0) \in
\mathfrak{R}$, that is, the truth values are extensions in
$\mathfrak{R}$.\\ The intensional semantics of the logic language
with the set of formulae $\L$ can be represented by the  mapping
\begin{center}
$~~~ \L ~\longrightarrow_I~ \D ~\Longrightarrow_{w \in \W}~
\mathfrak{R}$,
\end{center}
where $~\longrightarrow_I~$ is a \emph{fixed intensional}
interpretation $I:\L \rightarrow \D$ and $~\Longrightarrow_{w \in
\W}~$ is \emph{the set} of all extensionalization functions $h =
is(w):\D \rightarrow \mathfrak{R}$ in $\E$, where $is:\W \rightarrow
\E$ is the mapping from the set of possible worlds to the set of
 extensionalization functions.\\
 We define the mapping $I_n:\L_{op} \rightarrow
\mathfrak{R}^{\W}$, where $\L_{op}$ is the subset of formulae with
free variables (virtual predicates), such that for any virtual
predicate $\phi(x_1,...,x_k) \in \L_{op}$ the mapping
$I_n(\phi(x_1,...,x_k)):\W \rightarrow \mathfrak{R}$ is the
Montague's meaning (i.e., \emph{intension}) of this virtual
predicate \cite{Lewi86,Stal84,Mont70,Mont73,Mont74}, that is, the
mapping which returns with the extension of this (virtual) predicate
in every possible world in $\W$.
\end{definition}
We adopted this two-step intensional semantics, instead of well
known Montague's semantics (which lies in the construction of a
compositional and recursive semantics that covers both intension and
extension) because of a number of its weakness.\\
\textbf{Example 1}:  Let us consider the following two past
participles: 'bought' and 'sold'(with unary predicates $p_1^1(x)$,
'$x$ has been bought', and $p_2^1(x)$,'$x$ has been sold'). These
two different concepts in the Montague's semantics would have not
only the same extension but also their intension, from the fact that
their extensions are identical in every possible world. Within the
two-steps formalism we can avoid this problem by assigning two
different concepts (meanings) $u = I(p_1^1(x))$ and $ v =
I(p_2^1(x))$ in $\in D_1$. Notice that the same problem we have in
the Montague's semantics for two sentences with different meanings,
which bear the same truth value across all possible worlds: in the
Montague's semantics they will be forced to the \emph{same}
meaning.\\$\square$\\
 Another relevant question w.r.t. this two-step
interpretations of an intensional semantics is how in it is managed
the extensional identity relation $\doteq$ (binary predicate of the
identity) of the FOL. Here this extensional identity relation is
mapped into the binary concept $Id = I(\doteq(x,y)) \in D_2$, such
that $(\forall w \in \W)(is(w)(Id) = R_{=})$, where $\doteq(x,y)$
(i.e., $p_1^2(x,y)$) denotes an atom of the FOL of the binary
predicate for identity in FOL, usually written by FOL formula $x
\doteq y$ (here we prefer to distinguish this \emph{formal symbol}
$~ \doteq ~ \in P$ of the built-in identity binary predicate letter
in the FOL from the standard mathematical
symbol '$=$' used in all mathematical definitions in this paper).\\
 In what follows we will use the function $f_{<>}:\mathfrak{R}
\rightarrow \mathfrak{R}$, such that for any $R \in \mathfrak{R}$,
$f_{<>}(R) = \{<>\}$ if $R \neq \emptyset$; $\emptyset$ otherwise.
Let us define the following set of algebraic operators for
 relations in $\mathfrak{R}$:
\begin{enumerate}
\item binary operator $~\bowtie_{S}:\mathfrak{R} \times \mathfrak{R} \rightarrow
\mathfrak{R}$,
 such that for any two relations $R_1, R_2 \in
 \mathfrak{R}~$, the
 $~R_1 \bowtie_{S} R_2$ is equal
to the relation obtained by natural join
 of these two relations $~$ \verb"if"
 $S$ is a non empty
set of pairs of joined columns of respective relations (where the
first argument is the column index of the relation $R_1$ while the
second argument is the column index of the joined column of the
relation $R_2$); \verb"otherwise" it is equal to the cartesian
product $R_1\times R_2$. For example, the logic formula
$\phi(x_i,x_j,x_k,x_l,x_m) \wedge \psi (x_l,y_i,x_j,y_j)$ will be
traduced by the algebraic expression $~R_1 \bowtie_{S}R_2$ where
$R_1 \in \P(\D^5), R_2\in \P(\D^4)$ are the extensions for a given
Tarski's interpretation  of the virtual predicate $\phi, \psi$
relatively, so that $S = \{(4,1),(2,3)\}$ and the resulting relation
will have the following ordering of attributes:
$(x_i,x_j,x_k,x_l,x_m,y_i,y_j)$.
\item unary operator $~ \sim:\mathfrak{R} \rightarrow \mathfrak{R}$, such that for any k-ary (with $k \geq 0$)
relation $R \in  \P(\D^{k}) \subset \mathfrak{R}$
 we have that $~ \sim(R) = \D^k \backslash R \in \D^{k}$, where '$\backslash$' is the substraction of relations. For example, the
logic formula $\neg \phi(x_i,x_j,x_k,x_l,x_m)$ will be traduced by
the algebraic expression $~\D^5 \backslash R$ where $R$ is the
extensions for a given Tarski's interpretation  of the virtual
predicate $\phi$.
\item unary operator $~ \pi_{-m}:\mathfrak{R} \rightarrow \mathfrak{R}$, such that for any k-ary (with $k \geq 0$) relation $R \in \P(\D^{k}) \subset \mathfrak{R}$
we have that $~ \pi_{-m} (R)$ is equal to the relation obtained by
elimination of the m-th column of the relation $R~$ \verb"if" $1\leq
m \leq k$ and $k \geq 2$; equal to $~f_{<>}(R)~$ \verb"if" $m = k
=1$; \verb"otherwise" it is equal to $R$. For example, the logic
formula $(\exists x_k) \phi(x_i,x_j,x_k,x_l,x_m)$ will be traduced
by the algebraic expression $~\pi_{-3}(R)$ where $R$ is the
extensions for a given Tarski's interpretation  of the virtual
predicate $\phi$ and the resulting relation will have the following
ordering of attributes: $(x_i,x_j,x_l,x_m)$.
\end{enumerate}
Notice that the ordering of attributes of resulting relations
corresponds to the method used for generating the ordering of
variables in the tuples of free variables adopted for virtual
predicates.\\
  Analogously to Boolean algebras
 which are extensional models of propositional logic, we introduce an
 intensional algebra for this intensional FOL as follows.
\begin{definition}  \label{def:int-algebra} Intensional algebra for the intensional FOL  in Definition \ref{def:bealer} is a structure $~Alg_{int}
= ~(\D,  f, t, Id, Truth,  \{conj_{S}\}_{ S \in \P(\mathbb{N}^2)},
neg,
\\\{exists_{n}\}_{n \in \mathbb{N}})$,  $~~$  with
 binary operations  $~~conj_{S}:D_I\times D_I \rightarrow D_I$,
   unary operation  $~~neg:D_I\rightarrow D_I$,  unary
   operations $~~exists_{n}:D_{I}\rightarrow D_I$,  such that for any
extensionalization function $h \in \E$,
and $u \in D_k, v \in D_j$, $k,j \geq 0$,\\
1. $~h(Id) = R_=~$ and $~h(Truth) = \{<>\}$.\\
2. $~h(conj_{S}(u, v)) = h(u) \bowtie_{S}h(v)$, where $\bowtie_{S}$
is the natural join operation defined above and $conj_{S}(u, v) \in D_m$ where $m = k + j - |S|$
 if for every pair $(i_1,i_2) \in S$ it holds that $1\leq i_1 \leq k$, $1 \leq i_2 \leq j$ (otherwise $conj_{S}(u, v) \in D_{k+j}$).\\
3. $~h(neg(u)) = ~\sim(h(u)) = \D^k \backslash (h(u))$,
 where  $~\sim~$ is the operation
defined above and $neg(u) \in D_k$.\\
 4. $~h(exists_{n}(u)) =
\pi_{-n}(h(u))$, where $\pi_{-n}$ is the operation defined above and
$exists_n(u) \in D_{k-1}$ if $1 \leq n \leq k$ (otherwise $exists_n$
is the identity function).
\end{definition}
Notice that for $u \in D_0$, $~h(neg(u)) = ~\sim(h(u)) = \D^0
\backslash (h(u)) = \{<>\} \backslash (h(u)) \in \{f,t\}$.\\ We
define a derived operation $~~union:(\P(D_i)\backslash \emptyset)
\rightarrow D_i$, $i \geq 0$, such that, for any $B =
\{u_1,...,u_n\} \in \P(D_i)$ we have that
$union(\{u_1,...,u_n\})=_{def} ~u_1$ if $n = 1$; $
neg(conj_S(neg(u_1),conj_S(...,neg(u_n))...)$, where $S =
\{(l,l)~|~1 \leq l \leq i \}$, otherwise.\\ Than we obtain that for $n \geq 2$:\\
$h(union (B) = h(neg(conj_S(neg(u_1),conj_S(...,neg(u_n))...) \\
 = \D^i\backslash((\D^i\backslash h(u_1)
\bowtie_{S}...\bowtie_{S}(\D^i\backslash h(u_n)) \\
= \D^i\backslash((\D^i\backslash h(u_1)
\bigcap...\bigcap(\D^i\backslash h(u_n)) \\
= \bigcup\{ h(u_j)~|~1 \leq j \leq n\} =  ~\bigcup \{h(u) ~|~u \in
B\}$. \\
 Once one has found a method for specifying the interpretations of
singular terms of $\L$ (take in consideration the particularity of
abstracted terms), the Tarski-style definitions of truth and
validity for  $\L$ may be given in the customary way.
What is being south specifically is a method for characterizing the
intensional interpretations of singular terms of $\L$ in such a way
that a given singular abstracted term $\lessdot \phi
\gtrdot_{\alpha}^{\beta}$ will denote an appropriate property,
relation, or proposition, depending on the value of $m = |\alpha|$.
Thus, the mapping of intensional abstracts (terms)  into $\D$ we
will define differently from that given in the version of Bealer
\cite{Beal82}, as follows:
\begin{definition}  \label{def:abstraction} An intensional
interpretation $I$ can be extended to abstracted terms as follows:
for any abstracted term $\lessdot \phi \gtrdot_{\alpha}^{\beta}$  we
define that,\\
$I(\lessdot \phi \gtrdot_{\alpha}^{\beta} ) = union (\{I(\phi[\beta/
g(\beta)])~|~ g \in \D^{\overline{\beta}}\})$,\\
where $\overline{\beta}$ denotes the \verb"set" of elements in the
list $\beta$, and the assignments in  $\D^{\overline{\beta}}$ are
limited only to the variables in $\overline{\beta}$.
 \end{definition}
\textbf{Remark:} Here we can make the question if there is a sense
to extend the interpretation also to (abstracted) terms, because in
Tarski's interpretation of FOL we do not have any interpretation for
terms, but only the assignments for terms, as we defined previously
by the mapping $g^*:\T \rightarrow\D$. The answer is positive,
because the abstraction symbol $\lessdot \_~
\gtrdot_{\alpha}^{\beta}$ can be considered as a kind of the unary
built-in functional symbol of intensional FOL, so that we can apply
the Tarskian interpretation to this functional symbol into the fixed
mapping $I(\lessdot \_~ \gtrdot_{\alpha}^{\beta}):\L \rightarrow
\D$, so that for any $\phi \in \L$ we have that $I(\lessdot \phi
\gtrdot_{\alpha}^{\beta})$ is equal to the application of this
function to the value $\phi$, that is, to $I(\lessdot \_~
\gtrdot_{\alpha}^{\beta})(\phi)$. In such an approach we would
introduce also the typed variable $X$ for the formulae in $\L$, so
that the Tarskian assignment for this
functional symbol with variable $X$, with $g(X) = \phi \in \L$, can be given by:\\
$g^*(\lessdot \_~ \gtrdot_{\alpha}^{\beta}(X)) = I(\lessdot \_~
\gtrdot_{\alpha}^{\beta})(g(X)) = I(\lessdot \_~
\gtrdot_{\alpha}^{\beta})(\phi)\\ = ~<>\in D_{-1},~$ if
$~\overline{\alpha} \bigcup \overline{\beta}$ is not
equal to the set of free variables in $\phi$;\\
$= union (\{I(\phi[\beta/ g'(\beta)])~|~ g' \in
\D^{\overline{\beta}}\}) \in D_{|\overline{\alpha}|}$, otherwise.
\\$\square$\\
 Notice than if $\beta = \emptyset$ is the empty
list, then $I(\lessdot \phi \gtrdot_{\alpha}^{\beta} ) = I(\phi)$.
Consequently, the denotation of $\lessdot \phi\gtrdot $
 is equal to the meaning of a proposition $\phi$, that is, $~I(\lessdot \phi\gtrdot) =
I(\phi)\in D_0$.  In the case when $\phi$ is an atom
$p^m_i(x_1,..,x_m)$ then $I (\lessdot
p^m_i(x_1,..,x_m)\gtrdot_{x_1,..,x_m}) = I(p^m_i(x_1,..,x_m)) \in
D_m$, while $I (\lessdot p^m_i(x_1,..,x_m)\gtrdot^{x_1,..,x_m}) =
union (\{I(p^m_i(g(x_1),..,g(x_m)))~|~ g \in \D^{\{x_1,..,x_m\}} \})
\in D_0$,  with $h(I (\lessdot
p^m_i(x_1,..,x_m)\gtrdot^{x_1,..,x_m})) = h(I((\exists
x_1)...(\exists x_m)p^m_i(x_1,..,x_m))) \in \{f,t\}$.\\ For example,
$h(I(\lessdot p^1_i(x_1) \wedge \neg p^1_i(x_1) \gtrdot^{x_1})) =
h(I((\exists x_1)(\lessdot p^1_i(x_1) \wedge \neg p^1_i(x_1)
\gtrdot^{x_1}))) = f$.\\
The interpretation of a more complex
abstract $\lessdot \phi \gtrdot_\alpha^{\beta}$ is defined in terms
of the interpretations of the relevant syntactically simpler
expressions, because the interpretation of more complex formulae is
defined in terms of the interpretation of the relevant syntactically
simpler formulae, based on the intensional algebra above. For
example, $I(p_i^1(x) \wedge p_k^1(x)) =
conj_{\{(1,1)\}}(I(p_i^1(x)), I(p_k^1(x)))$, $I(\neg
\phi) = neg(I(\phi))$, $I(\exists x_i)\phi(x_i,x_j,x_i,x_k) = exists_3(I(\phi))$.\\
Consequently, based on the intensional algebra in Definition
\ref{def:int-algebra} and on intensional interpretations of
abstracted term in Definition \ref{def:abstraction}, it holds that
the interpretation of any formula in $\L$ (and any abstracted term)
will be reduced to an algebraic expression over interpretation of
primitive atoms in $\L$. This obtained expression is finite for any
finite formula (or abstracted term), and represents the \emph{
meaning} of such finite formula (or abstracted term).\\
The \emph{extension} of abstracted terms satisfy the following
property:
\begin{propo}  \label{prop:abstraction}
For any abstracted term $\lessdot \phi \gtrdot_{\alpha}^{\beta}$
with $|\alpha| \geq 1$ we have that $~~ h(I(\lessdot \phi
\gtrdot_{\alpha}^{\beta} )) = \pi_{- \beta}(h(I(\phi)))$,\\ where
$\pi_{- (y_1,...,y_k)} = \pi_{-y_1} \circ ...\circ \pi_{-y_1}$,
$\circ$ is the sequential composition of functions), and  $\pi_{-
\emptyset}$ is an identity.
\end{propo}
\textbf{Proof:} Let $\textbf{x}$ be a tuple of all free variables in
$\phi$, so that $\overline{\textbf{x}} = \overline{\alpha} \bigcup
\overline{\beta}$, $\alpha = (x_1,...,x_k)$, then we have that
$~~ h(I(\lessdot \phi \gtrdot_{\alpha}^{\beta} )) =\\
 = h(union (\{I(\phi[\beta/
g(\beta)])~|~ g \in \D^{\overline{\beta}}\}))$, from Def.
\ref{def:abstraction}\\
$= \bigcup \{h(I(\phi[\beta/ g(\beta)]))~|~ g \in
\D^{\overline{\beta}}\}$ \\
$= \bigcup \{ \{(g_1(x_1),...,g_1(x_k))~|~ g_1 \in
\D^{\overline{\alpha}}$ and $ h(I(\phi[\beta/ g(\beta)][\alpha/
g_1(\alpha)])) = t\}~|~ g \in \D^{\overline{\beta}}\}$\\
$=  \{g_1(\alpha)~|~ g_1 \in \D^{\overline{\alpha}\bigcup
\overline{\beta}}$ and $ h(I(\phi/g_1))
= t\}\\
=  \pi_{- \beta}(\{g_1(\textbf{x})~|~ g_1 \in
\D^{\overline{\textbf{x}}}$ and $ h(I(\phi/g_1))
= t\})\\
=  \pi_{- \beta}(\{g_1(\textbf{x})~|~ g_1 \in
\D^{\overline{\textbf{x}}}$ and $g_1(\textbf{x})\in h(I(\phi))
\})$, by (T) \\
$ =  \pi_{- \beta}(h(I(\phi)))$.
\\$\square$\\
 We can connect $\E$
 with a possible-world semantics. Such a correspondence is a natural identification of
 intensional logics with modal Kripke based logics.
\begin{definition} (Model): \label{def:Semant} A model for  intensional FOL with fixed
intensional interpretation $I$, which express the two-step
intensional semantics in Definition \ref{def:intensemant}, is the
Kripke structure $\M_{int} = (\W, \D, V)$, where $\W =
\{is^{-1}(h)~|~h \in \E\}$,  a mapping $~V:\W \times P \rightarrow
\bigcup_{n < \omega} \{t,f\}^{\D^n}$, with $P$ a set of predicate
symbols of the language, such that for any world $w = is^{-1}(h) \in
\W, p^n_i \in P$, and $(u_1,...,u_n) \in \D^n$ it holds that
$V(w,p^n_i)(u_1,...,u_n) = h(I(p^n_i(u_1,...,u_n)))$. The
satisfaction relation $\models_{w,g}$ for a given $w \in \W$ and
assignment $g \in \D{\V}$ is defined as follows:\\
1. $~{\M} \models_{w,g}p_i^k(x_1,...,x_k)~$ iff $~V(w,p_i^k)(g(x_1),...\\,g(x_k)) = t$,\\
 2. $~{\M} \models_{w,g} \varphi \wedge \phi~$ iff $~{\M} \models_{w,g} \varphi~$ and $~{\M} \models_{w,g}
 \phi$, \\
 3. $~{\M} \models_{w,g} \neg \varphi ~$ iff $~$ not ${\M} \models_{w,g}
\varphi$,\\
 4. $~\M \models_{w,g}
(\exists x) \phi ~~$ iff \\
4.1. $~\M \models_{w,g} \phi$, if $x$ is not a free
variable in $\phi$;\\
4.2. $~$exists $u \in \D$ such that $~\M \models_{w,g} \phi[x/u]$,
if $x$ is  a free variable in $\phi$.
 \end{definition}
It is easy to show that  the satisfaction relation $\models$ for
this Kripke semantics in a world $w = is^{-1}(h)$ is defined
by, $~~\M \models_{w,g} \phi~~$ iff $~~h(I(\phi/g)) = t$.\\
We can enrich this intensional FOL by another modal operators, as
for example the "necessity" universal operator $\Box$ with an
accessibility relation $~~\R =\W \times \W$, obtaining the S5 Kripke
structure $\M_{int} = (\W, \R, \D, V)$, in order to be able to
define the following equivalences between the abstracted terms
without free variables $\lessdot \phi \gtrdot_{\alpha}^{\beta_1}/g$
and $\lessdot \psi \gtrdot_{\alpha}^{\beta_2}/g$, where all free
variables (not in $\alpha$) are instantiated by $g \in \D^{\V}$
(here $A \equiv B$ denotes the formula $(A \Rightarrow B) \wedge (B
\Rightarrow A)$):
\begin{itemize}
  \item (Strong) intensional
equivalence (or \emph{equality}) "$\asymp$"  is defined by:\\
$\lessdot \phi \gtrdot_{\alpha}^{\beta_1}/g ~\asymp~  \lessdot \psi
\gtrdot_{\alpha}^{\beta_2}/g~~~~~$
iff $~~~~~\Box ( \phi[\beta_1/g(\beta_1)] \equiv \psi[\beta_2/g(\beta_2)])$, \\
 with $~{\M} \models_{w,g'}~\Box \varphi~$ iff $~$ for all $w'\in \W$, $(w,w')
\in {\R}$ implies ${\M} \models_{w',g'}\varphi$.\\
From Example 1 we have that $\lessdot p_1^1(x)\gtrdot_{x} ~\asymp~
\lessdot p_2^1(x) \gtrdot_{x}$, that is '$x$ has been bought' and
'$x$ has been sold' are intensionally equivalent, but they have not
the same meaning (the concept $I(p_1^1(x)) \in D_1$ is different
from $I(p_2^1(x)) \in D_1$).
  \item Weak intensional equivalence "$\approx$" is defined by:\\
$\lessdot \phi \gtrdot_{\alpha}^{\beta_1}/g ~\approx~  \lessdot \psi
\gtrdot_{\alpha}^{\beta_2}/g~~~~~$
iff $~~~~~\diamondsuit  \phi[\beta_1/g(\beta_1)] \equiv \diamondsuit \psi[\beta_2/g(\beta_2)]$.\\
The symbol  $\diamondsuit = \neg \Box \neg$ is the correspondent
existential modal operator.\\
This weak equivalence is used for P2P database integration in a
number of papers
\cite{Majk03s,Majk05w,Majk06J,Majk06Om,Majk08f,Majk08rdf,Majk08in}.
\end{itemize}
Notice that we do not use the intensional equality in  our language,
thus we do not need the correspondent operator in intensional
algebra $~Alg_{int}$ for the logic "necessity" modal operator
$\Box$.\\
 This
semantics is equivalent to the algebraic semantics for $\L$ in
\cite{Beal79} for the case of the conception where intensional
entities are considered to be \emph{equal} if and only if they are
\emph{necessarily equivalent}. Intensional equality is much stronger
that the standard \emph{extensional equality}  in the actual world,
just because requires the extensional equality in \emph{all}
possible worlds, in fact, if $\lessdot \phi
\gtrdot^{\beta_1}_{\alpha}/g \asymp \lessdot \psi
\gtrdot^{\beta_1}_{\alpha}/g~$ then $h(I(\lessdot A
\gtrdot^{\beta_1}_{\alpha}/g)) = h(I(\lessdot \psi
\gtrdot^{\beta_2}_{\alpha}/g))~$ for all extensionalization
functions $h \in \E$ (that is possible worlds $is^{-1}(h) \in
\widetilde{\W}$).\\
It is easy to verify that the intensional equality means that in
every possible world $w \in \widetilde{\W}$ the intensional entities
$u_1$ and $u_2$ have the same extensions.\\
Let the logic modal formula $\Box \phi[\beta_1/ g(\beta_1)]$, where
the assignment $g$ is applied only to free variables in $\beta_1$ of
a formula $\phi$ not in the list of variables in $\alpha =
(x_1,...,x_n)$, $n \geq 1$, represents a n-ary intensional concept
such that $I(\square \phi[\beta_1/ g(\beta_1)]) \in D_n$ and
$I(\phi[\beta_1/ g(\beta_1)]) = I(\lessdot \phi
\gtrdot^{\beta_1}_{\alpha}/g) \in D_n$. Then the extension of this
n-ary concept is equal to (here the mapping $necess:D_i\rightarrow
D_i$ for each $i \geq 0$
is a new operation of the intensional algebra $~Alg_{int}$ in Definition \ref{def:int-algebra}):\\
$ h(I(\Box \phi[\beta_1/ g(\beta_1)]) = h(necess(I(\phi[\beta_1/ g(\beta_1)]))) =\\
 = \{(g'(x_1),...,g'(x_n))~|~\M
\models_{w,g'} \square \phi[\beta_1/ g(\beta_1)]~$ and $g' \in \D^{\V}\}$\\
$ = \{ (g'(x_1),...,g'(x_n))~|~g' \in \D^{\V}$ and $\forall w_1
((w,w_1) \in \R$ implies $\M
\models_{w_1,g'}  \phi[\beta_1/ g(\beta_1)])~\}$\\
$= \bigcap_{h_1 \in ~\E} h_1(I(\phi[\beta_1/ g(\beta_1)]))$.\\
While,\\
$ h(I(\diamondsuit \phi[\beta_1/ g(\beta_1)]) =  h(I(\neg \Box \neg \phi[\beta_1/ g(\beta_1)]) \\
= h(neg(necess(I(\neg \phi[\beta_1/ g(\beta_1)])))) \\
= \D^n \backslash h(necess(I(\neg \phi[\beta_1/ g(\beta_1)]))) \\
= \D^n \backslash (\bigcap_{h_1 \in ~\E} h_1(I(\neg \phi[\beta_1/ g(\beta_1)]))) \\
= \D^n \backslash (\bigcap_{h_1 \in ~\E} h_1(neg(I(\phi[\beta_1/ g(\beta_1)])))) \\
= \D^n \backslash (\bigcap_{h_1 \in ~\E} \D^n \backslash
h_1(I(\phi[\beta_1/ g(\beta_1)])))\\
= \bigcup_{h_1 \in ~\E} h_1(I(\phi[\beta_1/ g(\beta_1)]))$.\\
 Consequently, the concepts $\Box \phi[\beta_1/ g(\beta_1)]$ and $\diamondsuit \phi[\beta_1/ g(\beta_1)]$
are the \emph{built-in} (or rigid) concept as well, whose extensions
does not depend on possible worlds.\\
Thus, two concepts are intensionally \emph{equal}, that is,
$\lessdot \phi \gtrdot_{\alpha}^{\beta_1}/g ~\asymp~  \lessdot \psi
\gtrdot_{\alpha}^{\beta_2}/g$, iff $~h(I(\phi[\beta_1/g(\beta_1)]))
=\\ h(I(\psi[\beta_2/g(\beta_2)]))$ for every $h$. Moreover, two
concepts are \emph{weakly }equivalent, that is, $\lessdot \phi
\gtrdot_{\alpha}^{\beta_1}/g ~\approx~  \lessdot \psi
\gtrdot_{\alpha}^{\beta_2}/g$, iff $~~h(I(\diamondsuit
\phi[\beta_1/g(\beta_1)])) = h(I(\diamondsuit
\psi[\beta_2/g(\beta_2)]))$.


\end{document}